\documentclass[twoside]{article}
\usepackage{qic}

\begin{document}
\setlength{\textheight}{8.0truein}    

\runninghead{The Solovay-Kitaev algorithm $\ldots$}
            {Christopher~M.~Dawson and Michael~A.~Nielsen$\ldots$}

\normalsize\textlineskip
\thispagestyle{empty}
\setcounter{page}{1}

\copyrightheading{0}{0}{2005}{000--000}

\vspace*{0.88truein}

\alphfootnote

\fpage{1}

\centerline{\bf
THE SOLOVAY-KITAEV ALGORITHM}
\vspace*{0.37truein}
\centerline{\footnotesize
CHRISTOPHER M. DAWSON}
\vspace*{0.015truein}
\centerline{\footnotesize\it School of Physical Sciences, The University
of Queensland, }
\baselineskip=10pt
\centerline{\footnotesize\it Brisbane, Queensland 4072, Australia}
\vspace*{10pt}
\centerline{\footnotesize
MICHAEL A. NIELSEN}
\vspace*{0.015truein}
\centerline{\footnotesize\it School of Physical Sciences, The University
of Queensland, }
\baselineskip=10pt
\centerline{\footnotesize\it Brisbane, Queensland 4072, Australia}
\vspace*{0.225truein}
\publisher{\today}

\vspace*{0.21truein}

\abstracts{
  This pedagogical review presents the proof of the Solovay-Kitaev
  theorem in the form of an efficient classical algorithm for
  compiling an arbitrary single-qubit gate into a sequence of gates
  from a fixed and finite set.  The algorithm can be used, for
  example, to compile Shor's algorithm, which uses rotations of $\pi /
  2^k$, into an efficient fault-tolerant form using only Hadamard,
  controlled-{\sc not}, and $\pi / 8$ gates.  The algorithm runs in
  $O(\log^{2.71}(1/\epsilon))$ time, and produces as output a sequence
  of $O(\log^{3.97}(1/\epsilon))$ quantum gates which is guaranteed to
  approximate the desired quantum gate to an accuracy within $\epsilon
  > 0$.  We also explain how the algorithm can be generalized to apply
  to multi-qubit gates and to gates from $SU(d)$.}{}{}

\vspace*{10pt}

\keywords{Solovay-Kitaev algorithm, universality, fault-tolerance}
\vspace*{3pt}

\vspace*{1pt}\textlineskip    

\section{Introduction}
\label{sect::introduction}

\begin{quote}
  ``I have been impressed by numerous instances of mathematical
  theories that are really about particular algorithms; these theories
  are typically formulated in mathematical terms that are much more
  cumbersome and less natural than the equivalent formulation today's
  computer scientists would use.'' --- Donald~E.~Knuth~\cite{Knuth74a}
\end{quote}

The Solovay-Kitaev (SK) theorem is one of the most important
fundamental results in the theory of quantum computation.  In its
simplest form the SK theorem shows that, roughly speaking, if a set of
single-qubit quantum gates generates a dense subset of $SU(2)$, then
that set is guaranteed to fill $SU(2)$ \emph{quickly}, i.e., it is
possible to obtain good approximations to any desired gate using
surprisingly short sequences of gates from the given generating set.

The SK theorem is important if one wishes to apply a wide variety of
different single-qubit gates during a quantum algorithm, but is
restricted to use gates from a more limited repertoire.  Such a
situation arises naturally in the context of fault-tolerant quantum
computation, where direct fault-tolerant constructions are typically
available only for a limited number of gates (e.g., the Clifford group
gates and $\pi / 8$ gate), but one may wish to implement a wider
variety of gates, such as the $\pi / 2^k$ rotations occurring in
Shor's algorithm~\cite{Shor94a,Shor97a}.  In this situation one must
use the limited set of fault-tolerant gates to build up accurate
fault-tolerant approximations to all the gates used in the algorithm,
preferably in the most efficient manner possible.  The SK theorem
tells us that such efficient constructions are possible.

The purpose of the present paper is to present the proof of the SK
theorem in the form of a concrete \emph{algorithm} which can be used
to approximate a desired unitary, $U$, using a given set of quantum
gates.  We believe that this algorithmic perspective is useful in its
own right, and, as the quote by Knuth suggests, also makes the ideas
behind the SK theorem considerably easier to understand than in
previous presentations.  However, we stress that the paper is a review
paper, as all the essential ideas appear in previous work; a detailed
history of the SK theorem and algorithm is given in
Section~\ref{sec:prior-work}.

The structure of the paper is as follows.
Section~\ref{sec:fundamentals} introduces the fundamental definitions
needed to understand the SK theorem and algorithm, and provides a
formal statement of the SK theorem.  Section~\ref{sec:sk-big-picture}
explains the SK algorithm for qubits, in the form of nine
easily-understood lines of pseudocode.  This section concentrates on
developing a broad understanding of how the SK algorithm works.  The
few remaining details of the algorithm which need verification are
explained in Section~\ref{sec:sk-verification}.
Section~\ref{sec:SK_qudits} extends the algorithm so it applies also
to \emph{qudits}, i.e., quantum systems with $d$-dimensional state
spaces.  As a special case we obtain a version of the SK algorithm
applicable to multiple qubits, simply by setting $d = 2^n$.
Section~\ref{sec:prior-work} discusses the relationship of the results
we have presented to prior work on the SK theorem.  This placement of
the discussion of prior work is perhaps somewhat unusual, and deserves
comment: we chose to defer the discussion until late in the paper in
order to make the comparisons between the present paper and prior work
as concrete and transparent as possible.  Section~\ref{sec:conclusion}
concludes.

\section{Fundamentals}
\label{sec:fundamentals}

In this section we introduce the fundamental definitions necessary to
understand the SK theorem and algorithm, and give a precise statement
of the SK theorem.

The basic goal of the SK algorithm is to take an arbitrary quantum
gate $U$ and find a good approximation to it using a sequence of gates
$g_1,\ldots,g_m$ drawn from some finite set $\mathcal{G}$.  In analogy
with classical computation, where an algorithm written in a
programming language such as C or Perl must be compiled into a
sequence of instructions in the native machine language, we will call
the set $\mathcal{G}$ of quantum gates an \emph{instruction set}, and
the process of finding good approximations \emph{compilation}.  Note
that while this nomenclature, introduced in~\cite{Harrow01a}, is
inspired by classical computer science, the analogies between the
classical and quantum concepts are obviously somewhat imprecise, and
should not be taken too seriously.

Let's define these concepts more precisely, beginning with instruction
sets:
\begin{definition}
  An \textbf{instruction set} $\mathcal{G}$ for a $d$-dimensional
  qudit is a \emph{finite} set of quantum gates satisfying:
\begin{enumerate}
\item All gates $g \in \mathcal{G}$ are in $SU(d)$, that is, they are
  unitary and have determinant $1$.

\item For each $g \in \mathcal{G}$ the inverse operation $g^\dagger$
  is also in $\mathcal{G}$.

\item $\mathcal{G}$ is a universal set for $SU(d)$, i.e., the group
  generated by $\mathcal{G}$ is dense in $SU(d)$.  This means that
  given any quantum gate $U \in SU(d)$ and any accuracy $\epsilon > 0$
  there exists a product $S \equiv g_1 \ldots g_m$ of gates from
  $\mathcal{G}$ which is an $\epsilon$-approximation to $U$.

\end{enumerate}
\end{definition}

The notion of approximation being used in this definition, and
throughout this paper, is approximation in operator norm.  That is, a
sequence of instructions generating a unitary operation $S$ is said to
be an \emph{$\epsilon$-approximation} to a quantum gate $U$ if $d(U,S)
\equiv \|U-S\| \equiv \sup_{\| \psi \| = 1} \|(U-S)\psi\| < \epsilon$.
In different contexts we will find it convenient to use both the
operator norm $\| \cdot \|$, and the associated distance function
$d(\cdot,\cdot)$. Another convenient notation is to use $S$ to refer
both to a sequence of instructions, $g_1, g_2, \ldots g_m$, and also
to the unitary operation $g_1 g_2 \ldots g_m$ corresponding to that
sequence.  With this convention the ``sequence'' $g_1,g_2,\ldots$
corresponds to the unitary formed by applying $g_1,g_2,\ldots$ in the
\emph{reverse} order.  While this is somewhat nonintuitive, in later
use it will prove to be the simplest and most natural convention.

We assume the reader is familiar with the elementary properties of the
operator norm, which we will mostly use without note.  One perhaps
somewhat less familiar property we'll have occasion to use is that if
$H$ is Hermitian then $d(I,\exp(iH)) = \max_E 2 \sin(|E|/2)$, where
the maximum is over all eigenvalues $E$ of $H$.  From this it follows
easily that $d(I,\exp(iH)) \leq \|H\|$.  Furthermore, provided all the
eigenvalues are in the range $-\pi$ to $\pi$ it follows that
$d(I,\exp(iH)) = \|H\|+O(\|H\|^3)$.

It is easy to understand the motivation behind parts~1 and~3 of our
definition of an instruction set.  Less clear is the reason for
part~2, namely, the requirement that if $g \in \mathcal{G}$ then we
must also have $g^\dagger \in \mathcal{G}$.  Our reason for imposing
this requirement is that it is used in the proof of the SK theorem,
not for any \emph{a priori} reason.  In particular, we will make use
of the easily-verified fact that if we have an instruction sequence
providing an $\epsilon$-approximation to some unitary $U$, then we can
obtain an $\epsilon$-approximation to $U^\dagger$ simply by reversing
the order of the sequence, and applying the corresponding inverse
instructions.  It appears to be an open problem whether a result
analogous to the SK theorem holds when condition~2 is relaxed.

The problem of quantum compilation may now be stated more precisely:
given an instruction set, $\mathcal{G}$, how may we approximate an
arbitrary quantum gate with a sequence of instructions from
$\mathcal{G}$; how does the sequence length increase as we require
more accurate approximations; and how quickly may we find such an
approximating sequence?  These final two questions are vitally
important, as an inaccurate or inefficient quantum compiler may negate
any improvements quantum computers have over classical computers.

As an illustration, suppose we have a quantum algorithm, such as
Shor's, which can be broken up into a sequence of $m$ quantum gates,
$U_1,\ldots,U_m$.  However, let us also suppose that not all of those
quantum gates are in our instruction set, and so it is necessary to
compile the algorithm so that each gate $U_j$ is re-expressed in terms
of the available instruction set.  If the compiled algorithm is
required to approximate the original to within $\epsilon$, then it is
sufficient that each gate $U_j$ be approximated to an accuracy
$\epsilon/m$~\cite{Bernstein97a}. If we make the \emph{a priori}
reasonable assumption~\cite{Preskill98b} that an accuracy of $\delta >
0$ requires an instruction sequence of length $O(1/\delta)$, then each
gate $U_j$ will require a sequence of length $O(m/\epsilon)$, and so
the total number of instructions required to implement the algorithm
is $O(m^2/\epsilon)$ --- a quadratic increase in the complexity of the
algorithm. For quantum algorithms such as Grover's search
algorithm~\cite{Grover96a,Grover97a}, which purports to give a
quadratic increase over the best classical algorithm, this may be
problematic if the available instruction set does not coincide with
the gates used in the algorithm.

The SK theorem may be stated as follows:
\begin{theorem}[Solovay-Kitaev]
  Let $\mathcal{G}$ be an instruction set for $SU(d)$, and let a
  desired accuracy $\epsilon > 0$ be given. There is a constant $c$
  such that for any $U\in SU(d)$ there exists a finite sequence $S$ of
  gates from $\mathcal{G}$ of length $O(\log^c{(1/\epsilon)})$ and
  such that $d(U,S) < \epsilon$.
\end{theorem}

Different proofs of the theorem give different values for $c$; the
proof we describe gives $c \approx 3.97$.  We discuss some other
proofs (and their corresponding values of $c$) in
Section~\ref{sec:prior-work}.

A critical issue not addressed in this statement of the SK theorem is
the question of how efficiently the approximating sequence $S$ may be
found on a classical computer.  The SK algorithm both provides a proof
of the SK theorem, and also shows that an approximating sequence $S$
may be found efficiently on a classical computer, with a running time
of $O(\log^{2.71}(1/\epsilon))$.

At first sight it appears paradoxical that the SK algorithm can
produce as output a gate sequence of length
$O(\log^{3.97}(1/\epsilon))$ in the asymptotically \emph{shorter} time
$O(\log^{2.71}(1/\epsilon))$.  We will see that the reason this is
possible is because the gate sequence has considerable redundancy,
enabling the output to be substantially compressed, i.e., not all the
gates need be explicitly output.  When this compression is not done,
the SK algorithm runs instead in time $O(\log^{3.97}(1/\epsilon))$.
Regardless of which method is used, the key point is that the running
time is \emph{polylogarithmic} in $1/\epsilon$.

Returning to our earlier example, suppose we have a quantum algorithm
expressed in terms of quantum gates $U_1,\ldots,U_m$.  If we wish for
the compiled algorithm to have accuracy $\epsilon$, then each gate
needs accuracy $\epsilon/m$.  Using the SK algorithm this can be
achieved using a sequence of $O(\log^{3.97}(m/\epsilon))$ quantum
gates, and a running time of $O(\log^{2.71}(m/\epsilon))$ on a
classical computer.  The total number of instructions required to
implement the algorithm is therefore $O(m \log^{3.97}(m/\epsilon))$,
with an associated classical compile time of $O(m
\log^{2.71}(m/\epsilon))$.  Thus, the SK algorithm reduces the
overhead due to quantum compilation from quadratic to polylogarithmic,
which is much more acceptable.

\section{The Solovay-Kitaev algorithm for qubits}
\label{sec:sk-big-picture}

In this section we present the main ideas used in the SK algorithm.
At the conclusion of the section the reader should have a good broad
understanding of how and why the algorithm works.  Rigorous
verification of a few details has been deferred until the next
section, in order that those details not obscure the big picture.  We
also restrict our attention in this section to that form of the SK
theorem which applies to \emph{qubits}.  The extension to
\emph{qudits} involves some new ideas, and is described in
Section~\ref{sec:SK_qudits}.

The SK algorithm may be expressed in nine lines of pseudocode.  We
explain each of these lines in detail below, but present it here in
its entirety both for the reader's reference, and to stress the
conceptual simplicity of the algorithm:
\\ \\
\begin{tt}
\indent \indent function Solovay-Kitaev(Gate $U$, depth $n$) \\
\indent \indent if ($n == 0$) \\
\indent \indent \indent Return Basic Approximation to $U$ \\
\indent \indent else \\
\indent \indent \indent Set $U_{n-1}$ = Solovay-Kitaev($U$,$n-1$) \\
\indent \indent \indent Set $V$, $W$ = GC-Decompose($U U_{n-1}^\dagger$) \\
\indent \indent \indent Set $V_{n-1}$ = Solovay-Kitaev($V$,$n-1$) \\
\indent \indent \indent Set $W_{n-1}$ = Solovay-Kitaev($W$,$n-1$) \\
\indent \indent \indent Return $U_n = V_{n-1} W_{n-1} V_{n-1}^\dagger
W_{n-1}^\dagger U_{n-1}$;
\end{tt}
\\

Let's examine each of these lines in detail.  The first line:
\\ \\
\begin{tt}
\indent \indent function Solovay-Kitaev(Gate $U$, depth $n$)
\end{tt}
\\ \\
indicates that the algorithm is a function with two inputs: an
arbitrary single-qubit quantum gate, $U$, which we desire to
approximate, and a non-negative integer, $n$, which controls the
accuracy of the approximation.  The function returns a sequence of
instructions which approximates $U$ to an accuracy $\epsilon_n$, where
$\epsilon_n$ is a decreasing function of $n$, so that as $n$ gets
larger, the accuracy gets better, with $\epsilon_n \rightarrow 0$ as
$n \rightarrow \infty$.  We describe $\epsilon_n$ in detail below.

The {\tt Solovay-Kitaev} function is recursive, so that to obtain an
$\epsilon_n$-approximation to $U$, it will call itself to obtain
$\epsilon_{n-1}$-approximations to certain unitaries.  The recursion
terminates at $n = 0$, beyond which no further recursive calls are
made:
\\ \\
\begin{tt}
\indent \indent if ($n == 0$) \\
\indent \indent \indent Return Basic Approximation to $U$
\end{tt}
\\ \\
In order to implement this step we assume that a preprocessing stage
has been completed which allows us to find a basic
$\epsilon_0$-approximation to arbitrary $U \in SU(2)$.  Since
$\epsilon_0$ is a constant, in principle this preprocessing stage may
be accomplished simply by enumerating and storing a large number of
instruction sequences from $\mathcal{G}$, say up to some sufficiently
large (but fixed) length $l_0$, and then providing a lookup routine
which, given $U$, returns the closest sequence.  Appropriate values
for $\epsilon_0$ and $l_0$ will be discussed later in this section.

At higher levels of recursion, to find an $\epsilon_n$-approximation
to $U$, we begin by finding an $\epsilon_{n-1}$-approximation to $U$:
\\ \\
\begin{tt}
\indent \indent else \\
\indent \indent \indent Set $U_{n-1}$ = Solovay-Kitaev($U$,$n-1$)
\end{tt}
\\ \\
We will use $U_{n-1}$ as a step towards finding an improved
approximation to $U$.  Defining $\Delta \equiv UU_{n-1}^\dagger$, the
next three steps of the algorithm aim to find an
$\epsilon_n$-approximation to $\Delta$, where $\epsilon_n$ is some
improved level of accuracy, i.e., $\epsilon_n < \epsilon_{n-1}$.
Finding such an approximation also enables us to obtain an
$\epsilon_n$-approximation to $U$, simply by concatenating our exact
sequence of instructions for $U_{n-1}$ with our
$\epsilon_n$-approximating sequence for $\Delta$.

How do we find such an approximation to $\Delta$?  First, observe that
$\Delta$ is within a distance $\epsilon_{n-1}$ of the identity.  This
follows from the definition of $\Delta$ and the fact that $U_{n-1}$ is
within a distance $\epsilon_{n-1}$ of $U$.

Second, decompose $\Delta$ as a group commutator $\Delta = VWV^\dagger
W^\dagger$ of unitary gates $V$ and $W$.  For any $\Delta$ it turns
out --- this is not obvious --- that there is always an infinite set
of choices for $V$ and $W$ such that $\Delta = VWV^\dagger W^\dagger$.
For our purposes it is important that we find $V$ and $W$ such that
$d(I,V), d(I,W) < c_{\rm gc} \sqrt{\epsilon_{n-1}}$ for some constant
$c_{\rm gc}$.  We call such a decomposition a \emph{balanced group
  commutator}.  In Subsection~\ref{subsec:balanced-su2} we will use
the fact that $d(I,\Delta) < \epsilon_{n-1}$ to show that such a
balanced group commutator can always be found:
\\ \\
\begin{tt}
  \indent \indent \indent Set $V$, $W$ = GC-Decompose($U U_{n-1}^\dagger$)
\end{tt}
\\ \\
For practical implementations we will see below that it is useful to
have $c_{\rm gc}$ as small as possible; the arguments of
Subsection~\ref{subsec:balanced-su2} show that $c_{\rm gc} \approx
1/\sqrt{2}$.

The next step is to find instruction sequences which are
$\epsilon_{n-1}$-approximations to $V$ and $W$:
\\ \\
\begin{tt}
\indent \indent \indent Set $V_{n-1}$ = Solovay-Kitaev($V$,$n-1$) \\
\indent \indent \indent Set $W_{n-1}$ = Solovay-Kitaev($W$,$n-1$)
\end{tt}
\\ \\
Remarkably, in Subsection~\ref{subsec:approximating_commutators} we
show that the group commutator of $V_{n-1}$ and $W_{n-1}$ turns out to
be an $\epsilon_n \equiv c_{\rm approx}
\epsilon_{n-1}^{3/2}$-approximation to $\Delta$, for some small
constant $c_{\rm approx}$. Provided $\epsilon_{n-1} < 1/c_{\rm
  approx}^2$, we see that $\epsilon_n < \epsilon_{n-1}$, and this
procedure therefore provides an \emph{improved} approximation to
$\Delta$, and thus to $U$.  This is surprising, since we are using
imperfect approximations to $V$ and $W$ to obtain an improved
approximation to the group commutator $VWV^\dagger W^\dagger$.  We
will see in Subsection~\ref{subsec:approximating_commutators} that the
reason this improved approximation is possible is because of
cancellation of error terms in the group commutator.


The constant $c_{\rm approx}$ is important as it determines the
precision $\epsilon_0$ required of the initial approximations. In
particular, we see that for this construction to guarantee that
$\epsilon_0 > \epsilon_1 > \ldots$ we must have $\epsilon_0 <
1/c^2_{\rm approx}$. Following the discussion in
Subsection~\ref{subsec:approximating_commutators}, we obtain $c_{\rm
  approx} \approx 8 c_{\rm gc} \approx 4\sqrt{2}$ and therefore
require $\epsilon_0 < 1/32$.  Of course, the analysis in
Subsection~\ref{subsec:approximating_commutators} simply bounds
$c_{\rm approx}$, and it is possible that a more detailed analysis
will give better bounds on $c_{\rm approx}$, and thus on $\epsilon_0$.
Numerically we have found that for the single-qubit instruction set
consisting of the Hadamard gate, the $\pi / 8$ gate, and its inverse,
$\epsilon_0 \approx 0.14$ and $l_0 = 16$ is sufficient for practical
purposes.

The algorithm concludes by returning the sequences approximating the
group commutator, as well as $U_{n-1}$:
\\ \\
\begin{tt}
  \indent \indent \indent Return $U_n = V_{n-1} W_{n-1} V_{n-1}^\dagger
W_{n-1}^\dagger U_{n-1}$;
\end{tt}
\\ \\
Summing up, the function {\tt Solovay-Kitaev(U, n)} returns a sequence
which provides an $\epsilon_n = c_{\rm approx}
\epsilon_{n-1}^{3/2}$-approximation to the desired unitary $U$.  The
five constituents in this sequence are all obtained by calling the
function at the $n-1$th level of recursion.

\textbf{Analysis:} We now analyse the runtime of the SK algorithm, and
the accuracy of the approximating sequence of instructions.  Let $l_n$
be the length of the sequence of instructions returned by {\tt
  Solovay-Kitaev(U, n)}, and let $t_n$ be the corresponding runtime.
Inspection of the pseudocode and the above discussion shows that we
have the recurrences
\begin{eqnarray}
  \epsilon_n & = & c_{\rm approx} \epsilon_{n-1}^{3/2} \\
  l_n & = & 5 l_{n-1} \\
  t_n & \leq & 3 t_{n-1}+{\rm const},
\end{eqnarray}
where the constant overhead in $t_n$ comes from the need to perform
tasks like finding the balanced group commutator, and other small
overheads in the pseudocode.  Note that the cost of doing $t_n$ is
assumed to come entirely from the calls to {\tt Solovay-Kitaev(.,
  n-1)}, and we have assigned a constant cost for the final line of
the algorithm, which returns the new sequence $U_n$.  This assumption
is justified if the final line is implemented using pointers to the
output from the earlier calls to {\tt Solovay-Kitaev(., n-1)}, rather
than returning an actual sequence of gates.  This enables us to avoid
explicitly returning the (redundant) sequences for $V_{n-1}^\dagger$
and $W_{n-1}^\dagger$.  If this is not done, then this analysis needs
a little modification; we won't go through the details, but the end
result is that the time and length both scale as
$O(\log^{3.97}(1/\epsilon))$.  Assuming that pointers have been used,
the recurrences above imply:
\begin{eqnarray}
  \epsilon_n & = & \frac{1}{c_{\rm approx}^2}
  \left(\epsilon_0 c_{\rm approx}^2 \right)^{\left(\frac{3}{2}\right)^n}. \\
  l_n & = & O(5^n) \label{eq:inter-length} \\
  t_n & = & O(3^n). \label{eq:inter-time}
\end{eqnarray}
To obtain a given accuracy $\epsilon$ it follows that we must choose
$n$ to satisfy the equation:
\begin{eqnarray}
  n = \left\lceil \frac{ \ln \left[ \frac{\ln(1/\epsilon c_{\rm approx}^2)}
      {\ln(1/\epsilon_0 c_{\rm approx}^2)} \right]}
  { \ln(3/2) } \right\rceil.
\end{eqnarray}
Substituting this value of $n$ back into Eqs.~(\ref{eq:inter-length})
and~(\ref{eq:inter-time}) we obtain expressions for the length of the
approximating sequence and the time of execution, now as a function of
$\epsilon$ rather than $n$:
\begin{eqnarray}
  l_\epsilon & = & O\left( \ln^{\ln 5/ \ln(3/2)} (1/\epsilon) \right) \\
  t_\epsilon & = & O\left( \ln^{\ln 3/ \ln(3/2)} (1/\epsilon) \right).
\end{eqnarray}
These are the desired expressions for the length and execution time.
Note that the exponent in the first expression is $\ln 5 / \ln(3/2)
\approx 3.97$, while the exponent in the second expression is $\ln 3 /
\ln(3/2) \approx 2.71$.

A caveat to this discussion is that we have entirely ignored the
\emph{precision} with which arithmetical operations are carried out.
In practice we cannot compute with precisely specified unitary
operations, since such operations in general require an infinite
amount of classical information to specify.  Instead, we must work
with approximations to such operations, and do all our calculations
using finite precision arithmetic.  A complete analysis of the SK
algorithm should include an account of the cost of such
approximations.  In practice, we have found that for the values of
precision of most interest to us, standard floating-point arithmetic
works admirably well, and there seems to be little need for such an
analysis.

\section{Verification of details}
\label{sec:sk-verification}

In this section we verify those details of the SK algorithm not fully
explained in the previous section.
Subsection~\ref{subsec:balanced-su2} describes how to find balanced
group commutators generating a desired unitary, while
Subsection~\ref{subsec:approximating_commutators} discusses the error
in the group commutator $VWV^\dagger W^\dagger$ when only
approximations to the constituent unitaries are available.

\subsection{Balanced commutators in $SU(2)$}
\label{subsec:balanced-su2}

%
%
The earlier discussion of balanced group commutators was in a somewhat
specialized notation, which arose out of the application to the
Solovay-Kitaev algorithm.  To emphasize the generality of the results
in this subsection we will use a more generic notation.  In
particular, suppose $U \in SU(2)$ satisfies $d(I,U) < \epsilon$.  Our
goal is to find $V$ and $W$ so that the group commutator satisfies
$VWV^\dagger W^\dagger = U$, and such that $d(I,V), d(I,W) < c_{\rm
  gc} \sqrt{\epsilon}$ for some constant $c_{\rm gc}$.

In order to find such a $V$ and $W$ we first examine what seems like a
special case.  In particular, we choose $V$ to be a rotation by an
angle $\phi$ about the $\hat x$ axis of the Bloch sphere, and $W$ to
be a rotation by an angle $\phi$ about the $\hat y$ axis of the Bloch
sphere.  The resulting group commutator $VWV^\dagger W^\dagger$ is a
rotation about some axis $\hat n$ on the Bloch sphere, by an angle
$\theta$ satisfying:
\begin{eqnarray} \label{eq:angle}
  \sin(\theta/2) = 2 \sin^2(\phi/2) \sqrt{1-\sin^4(\phi/2)}.
\end{eqnarray}
There are many ways of verifying Eq.~(\ref{eq:angle}), including
simple brute force calculation, perhaps with the aid of a computer
mathematics package.  One somewhat more elegant way of verifying
Eq.~(\ref{eq:angle}) is as follows.  First, observe that $V^\dagger$
is a rotation by an angle $\phi$ about the $-\hat x$ axis, and thus $W
V^\dagger W^\dagger$ must be a rotation by an angle $\phi$ about the
axis $\hat m$ that results when $-\hat x$ is rotated by an angle
$\phi$ about the $\hat y$ axis.  That is, $\hat m (-\cos(\phi),0,\sin(\phi))$.  The commutator $V (WV^\dagger
W^\dagger)$ is thus the composition of these two rotations.  The
resulting angle of rotation is easy to calculate (see, e.g.,
Exercise~4.15 on page~177 of~\cite{Nielsen00a}), with
Eq.~(\ref{eq:angle}) the result after some simplification. The exact
form of the axis of rotation, $\hat n$, is not so important for our
purposes, but it is easily computed using similar techniques.

We can easily invert this construction.  Suppose now that $U$ is a
rotation by an arbitrary angle $\theta$ about an arbitrary axis $\hat
p$ on the Bloch sphere.  Define $\phi$ to be a solution to
Eq.~(\ref{eq:angle}) for the given value of $\theta$, and define $V$
and $W$ to be rotations by $\phi$ about the $\hat x$ and $\hat y$ axes
of the Bloch sphere.  Then it is easy to see that $U$ must be
conjugate to a rotation by $\theta$ about the $\hat n$ axis identified
in the previous two paragraphs, i.e., $U = S (VWV^\dagger W^\dagger)
S^\dagger$, for some easily-computed unitary $S$.  It follows that
\begin{eqnarray}
  U = \tilde V \tilde W \tilde V^\dagger \tilde W^\dagger,
\end{eqnarray}
where $\tilde V \equiv S V S^\dagger$ and $\tilde W \equiv S W
S^\dagger$.  Thus, we can express any unitary $U$ rotating the Bloch
sphere by an angle $\theta$ as the group commutator of unitaries $V$
and $W$ which rotate the Bloch sphere by an angle $\phi$, providing
$\theta$ and $\phi$ are related by Eq.~(\ref{eq:angle}).

To conclude the argument, note that for a unitary $T$ rotating the
Bloch sphere by an angle $\tau$, the distance to the identity
satisfies $d(I,T) = 2 \sin(\tau/4) = \tau/2 + O(\tau^3)$.  Combining
this observation with Eq.~(\ref{eq:angle}), we see that for $U$ near
the identity we can express $U$ in terms of a group commutator of $V$
and $W$ satisfying $d(I,U) \approx 2 d(I,V)^2 = 2 d(I,W)^2$.  That is,
we have:
\begin{eqnarray}
  U & = & VWV^\dagger W^\dagger \\
  d(I,V) = d(I,W) & \approx & \sqrt{\frac{d(I,U)}{2}} <
\sqrt{\frac{\epsilon}{2}}.
\end{eqnarray}
This is the desired balanced group commutator, and gives $c_{\rm gc}
\approx 1/\sqrt{2}$.  This argument can also easily be modified to
give a more rigorous bound on $c_{\rm gc}$.  The details are tedious,
but easy to supply, and so we won't do this here.

\subsection{Approximating a commutator}
\label{subsec:approximating_commutators}

Just as in the last subsection, for clarity we state the main result
of this subsection in general terms, rather than in the special
notation used in Section~\ref{sec:sk-big-picture}.

\begin{lemma} {} \label{lemma:approx-comm}
  Suppose $V,W,\tilde V,$ and $\tilde W$ are unitaries such that
  $d(V,\tilde V),d(W,\tilde W) < \Delta$, and also $d(I,V), d(I,W) <
  \delta$.  Then:
\begin{eqnarray}
  d(VWV^\dagger W^\dagger, \tilde V \tilde W \tilde V^\dagger
  \tilde W^\dagger) < 8 \Delta \delta + 4\Delta \delta^2 + 8 \Delta^2 +
  4 \Delta^3 + \Delta^4.
\end{eqnarray}
\end{lemma}

In applying this lemma to the SK algorithm, we replace $\Delta$ by
$\epsilon_{n-1}$, and $\delta$ by $c_{\rm gc} \sqrt{\epsilon_{n-1}}$.
This gives rise to the inequality:
\begin{eqnarray}
  d(VWV^\dagger W^\dagger, \tilde V \tilde W \tilde V^\dagger
  \tilde W^\dagger) < c_{\rm approx} \epsilon_{n-1}^{3/2},
\end{eqnarray}
where $c_{\rm approx} \approx 8 c_{\rm gc}$.  With a little more
detailed work we can modify this argument to give a rigorous upper
bound on $c_{\rm approx}$.  However, just as in the last subsection,
the details of this argument are tedious and not especially
enlightening, but easy to supply, and so we won't do this here.

\textbf{Proof:} We begin by writing $\tilde V = V+\Delta_V, \tilde W W + \Delta_W$, which gives:
\begin{eqnarray}
  \tilde V \tilde W \tilde V^\dagger \tilde W^\dagger & = &
  VWV^\dagger W^\dagger
  + \Delta_V W V^\dagger W^\dagger + V \Delta _W V^\dagger W^\dagger
  + V W \Delta_V^\dagger W^\dagger + VWV^\dagger \Delta_W^\dagger \nonumber \\
  & & + O(\Delta^2) + O(\Delta^3) + O(\Delta^4). \label{eq:expansion}
\end{eqnarray}
The $O(\Delta^2)$ terms are terms like $\Delta_V \Delta_W V^\dagger
W^\dagger$.  There are ${4 \choose 2} = 6$ such terms. The
$O(\Delta^3)$ terms are terms like $\Delta_V \Delta_W \Delta_V^\dagger
W^\dagger$, of which there are ${4 \choose 3} = 4$ terms.  There is
just a single $O(\Delta^4)$ term, $\Delta_V \Delta_W \Delta_V^\dagger
\Delta_W^\dagger$.  It follows from these observations,
Eq.~(\ref{eq:expansion}), and the triangle inequality that:
\begin{eqnarray}
  d(VWV^\dagger W^\dagger, \tilde V \tilde W \tilde V^\dagger
  \tilde W^\dagger) & < &
  \| \Delta_V W V^\dagger W^\dagger + V \Delta _W V^\dagger W^\dagger
  + V W \Delta_V^\dagger W^\dagger + VWV^\dagger \Delta_W^\dagger \|
  \nonumber \\
  & & + 6 \Delta^2 + 4 \Delta^3 + \Delta^4.
\end{eqnarray}
To complete the proof it suffices to prove that $\| \Delta_V W
V^\dagger W^\dagger + V W \Delta_V^\dagger W^\dagger \| < \Delta^2 + 4
\Delta \delta + 2\Delta \delta^2$ and $\| V \Delta _W V^\dagger
W^\dagger + VWV^\dagger \Delta_W^\dagger \| < \Delta^2 + 4 \Delta
\delta + 2\Delta \delta^2$.  The proofs of the two results are
analogous, with the roles of $V$ and $W$ interchanged, and so we only
provide the details of the first proof.  We expand $W = I+\delta_W$,
so that
\begin{eqnarray}
\Delta_V W V^\dagger W^\dagger + V W \Delta_V^\dagger W^\dagger  \Delta_V V^\dagger + V\Delta_V^\dagger + O(\Delta \delta) + O(\Delta \delta^2).
\end{eqnarray}
In this expression the $O(\Delta \delta)$ terms are terms like
$\Delta_V \delta_W V^\dagger W^\dagger$.  By inspection we see that
there are four such terms.  The $O(\Delta \delta^2)$ terms are terms
like $\Delta V \delta_W V^\dagger \delta_W^\dagger$, and, again by
inspection, we see that there are two such terms.  It follows that:
\begin{eqnarray}
\| \Delta_V W V^\dagger W^\dagger + V W \Delta_V^\dagger W^\dagger \| <
 \| \Delta_V V^\dagger + V\Delta_V^\dagger \| + 4 \Delta \delta +
 2\Delta \delta^2.
\end{eqnarray}
The unitarity of $V$ and $V+\Delta_V$ implies that $\Delta_V V^\dagger
+ V \Delta_V^\dagger = -\Delta_V \Delta_V^\dagger$.  Combining these
observations and using the triangle inequality we obtain
\begin{eqnarray}
\| \Delta_V W V^\dagger W^\dagger + V W \Delta_V^\dagger W^\dagger \| <
 \Delta^2 + 4 \Delta \delta + 2\Delta \delta^2,
\end{eqnarray}
which completes the proof of the lemma.

\section{The Solovay-Kitaev algorithm for qudits}
\label{sec:SK_qudits}

In this section we present a generalization of the SK algorithm that
can be applied to qudits, i.e., to $d \times d$ unitary gates.  The
main difference is in the group commutator decomposition.  In
Subsection~\ref{subsec:modified_algorithm} we give a broad description
of the modified algorithm, and in Subsection~\ref{subsec:balanced-sud}
we describe the modified group commutator decomposition.

\subsection{The modified algorithm}
\label{subsec:modified_algorithm}

The modified pseudocode for the qudit SK algorithm is as follows:
\\ \\
\begin{tt}
\indent \indent function Solovay-Kitaev(Gate $U$, depth $n$) \\
\indent \indent if ($n == 0$) \\
\indent \indent \indent Return Basic Approximation to $U$ \\
\indent \indent else \\
\indent \indent \indent Set $U_{n-1}$ = Solovay-Kitaev($U$,$n-1$) \\
\indent \indent \indent Set $V$, $W$ = GC-Approx-Decompose($U U_{n-1}^\dagger$) \\
\indent \indent \indent Set $V_{n-1}$ = Solovay-Kitaev($V$,$n-1$) \\
\indent \indent \indent Set $W_{n-1}$ = Solovay-Kitaev($W$,$n-1$) \\
\indent \indent \indent Return $U_n = V_{n-1} W_{n-1} V_{n-1}^\dagger
  W_{n-1}^\dagger U_{n-1}$;
\\
\end{tt}

Comparing with the earlier pseudocode for the qubit SK algorithm, we
see that the only explicit difference is in the line taking the group
commutator:
\\ \\
\begin{tt}
\indent \indent \indent Set $V$, $W$ = GC-Approx-Decompose($U U_{n-1}^\dagger$)
\end{tt}
\\ \\
Recall that in the qubit algorithm the corresponding line finds a
balanced group commutator decomposition, i.e., finds $V$ and $W$ such
that (a) $VWV^\dagger W^\dagger = \Delta \equiv U U_{n-1}^\dagger$,
and (b) $d(I,V), d(I,W) < c_{\rm gc} \sqrt{\epsilon_{n-1}}$ for some
constant $c_{\rm gc}$.

In the qudit algorithm this line finds a balanced group commutator
which \emph{approximates} $\Delta$.  More precisely, we find $V$ and
$W$ such that (a) $d(VWV^\dagger W^\dagger,\Delta) < c_{\rm gc'}
\epsilon_{n-1}^{3/2}$, for some constant $c_{\rm gc'}$, and (b)
$d(I,V), d(I,W) < c_{\rm gc''} \sqrt{\epsilon_{n-1}}$ for some
constant $c_{\rm gc''}$.  The explicit procedure for doing this is
explained in Subsection~\ref{subsec:balanced-sud}, which shows that
$c_{\rm gc'} \approx 4 d^{3/4} ((d-1)/2)^{3/2}$ and $c_{\rm gc''}
\approx d^{1/4} ((d-1)/2)^{1/2}$, where $d$ is the dimensionality of
the Hilbert space we are working in.

The remaining lines work just as in the qubit algorithm, finding
instruction sequences $V_{n-1}$ and $W_{n-1}$ which are
$\epsilon_{n-1}$-approximations to $\Delta$, and then returning the
group commutator corresponding to those sequences, together with the
sequence for $U_{n-1}$, as the output of the algorithm:
\\ \\
\begin{tt}
\indent \indent \indent Set $V_{n-1}$ = Solovay-Kitaev($V$,$n-1$) \\
\indent \indent \indent Set $W_{n-1}$ = Solovay-Kitaev($W$,$n-1$) \\
\indent \indent \indent Return $U_n = V_{n-1} W_{n-1} V_{n-1}^\dagger
W_{n-1}^\dagger U_{n-1}$;
\end{tt}
\\ \\
Although the steps are the same, the analysis is a little different,
due to the fact that the group commutator of $V$ and $W$ only
approximates $\Delta$.  In particular, we have
\begin{eqnarray}
  d(V_{n-1} W_{n-1} V_{n-1}^\dagger W_{n-1}^\dagger, \Delta) & \leq &
  d(V_{n-1} W_{n-1} V_{n-1}^\dagger W_{n-1}^\dagger, VWV^\dagger W^\dagger) +
  d(VWV^\dagger W^\dagger,\Delta) \nonumber \\
  & & \\
  & < & \left( c_{\rm gc'} +c_{\rm approx} \right) \epsilon_{n-1}^{3/2},
\end{eqnarray}
where $c_{\rm approx}$ is the same constant that arose in the qubit
algorithm (Subsection~\ref{subsec:approximating_commutators}), where
we estimated $c_{\rm approx} \approx 8 c_{\rm gc''}$, in the notation
of the current section.  Thus, just as for the qubit algorithm, the
qudit algorithm returns a sequence $U_n$ which provides an $\epsilon_n
= O(\epsilon_{n-1}^{3/2})$- approximation to the desired unitary, $U$.
Furthermore, the five constituents in the sequence are all obtained by
calling the function at the $n-1$th level of recursion, just as in the
qubit algorithm.

The analysis of the accuracy and running time of the qudit SK
algorithm proceeds in a fashion analogous to the qubit algorithm,
giving $l_\epsilon = O\left( \ln^{\ln 5/ \ln(3/2)} (1/\epsilon)
\right)$ and $t_\epsilon = O\left( \ln^{\ln 3/ \ln(3/2)} (1/\epsilon)
\right)$.  An important practical caveat concerns the difficulty of
constructing the lookup table used to obtain the basic
$\epsilon_0$-approximations.  This is done by enumerating all possible
gate sequences up to some sufficient length.  $SU(d)$ is a manifold of
dimension $d^2 -1$, so if we wish to approximate every gate in $SU(d)$
to within $\epsilon_0$ then we generate $O(1/\epsilon_0^{d^2 -1})$
sequences. For an instruction set $\mathcal{G}$ there are
$O(|\mathcal{G}|^l)$ sequences of length $\leq l$, some fraction of
which may be redundant.  We will therefore need to enumerate all
sequences up to a length $l_0$ satisfying
\begin{equation}
l_0 \geq O\left(\frac{d^2 - 1}{\log |\mathcal{G}|} \log(1/\epsilon_0)\right) .
\end{equation}
The complexity of the enumeration is exponential in $l_0$ and thus
scales quite poorly with $d$.  While our algorithm is practical for
the small values of $d$ of most interest in applications to
fault-tolerance (e.g., $d=2,3,4$), it will require great computational
effort to scale it to substantially larger values of $d$.

\subsection{Balanced commutators in $SU(d)$}
\label{subsec:balanced-sud}

In this subsection our goal is to show that if $U$ satisfies $d(I,U) <
\epsilon$ then there exist $V$ and $W$ satisfying $d(VWV^\dagger
W^\dagger,U) < c_{\rm gc'} \epsilon^{3/2}$ and such that $d(I,V),
d(I,W) < c_{\rm gc''} \sqrt{\epsilon}$, for some constants $c_{\rm
  gc'}$ and $c_{\rm gc''}$.  The proof combines two lemmas.

\begin{lemma} {} \label{lemma:lie-solution} (Based on Theorem~4.5.2 on page~288
  of~\cite{Horn91a}) Let $H$ be a traceless $d$-dimensional Hermitian
  matrix.  Then we can find Hermitian $F$ and $G$ such that:
\begin{eqnarray}
  [F,G]          & = & iH \\
  \label{eq:norm_commutator_inequalities}
  \| F \|, \| G \| & \leq & d^{1/4} \left( \frac{d-1}{2} \right)^{1/2}
  \sqrt{\|H\|}.
\end{eqnarray}
\end{lemma}

A variant of this lemma may be found in Problem~8.15 on page~79
of~\cite{Kitaev02a}.  The variant in~\cite{Kitaev02a} has the
advantage that the prefactor to $\sqrt{\|H\|}$ on the right-hand side
of Eq.~(\ref{eq:norm_commutator_inequalities}) is replaced by a
constant that does not depend on $d$.  We use the present version
since the proof is a little simpler, and it suffices to establish the
correctness of the SK algorithm.  Readers serious about optimizing
this aspect of the SK algorithm should consult~\cite{Kitaev02a} for
details.

\textbf{Proof:} It is convenient to work in a basis which is Fourier
conjugate to the basis in which $H$ is diagonal.  (This can, of
course, be done, since the commutator bracket is preserved under
conjugation, i.e., $S [ A, B ] S^\dagger = [S A S^\dagger,S B
S^\dagger]$ for any matrices $A$ and $B$, and any unitary $S$.) In
this basis $H$ has the representation
\begin{eqnarray} \label{eq:fourier-H}
  H = W \mbox{diag}(E_1,\ldots,E_d) W^\dagger,
\end{eqnarray}
where $E_1,\ldots,E_d$ are the eigenvalues of $H$, and $W$ is the
Fourier matrix, with elements $W_{jk} \equiv \omega^{jk}/\sqrt{d}$,
where $\omega \equiv \exp(2\pi i /d)$ is a $d$th root of unity.  From
Eq.~(\ref{eq:fourier-H}) we see that in this basis the diagonal matrix
elements of $H$ vanish:
\begin{eqnarray}
  H_{jj} = \frac{\sum_k \omega^{jk} E_k {\omega^{jk}}^*}{d}
 = \frac{\sum_k E_k}{d}
 = \frac{\mbox{tr}(H)}{d} =  0.
\end{eqnarray}
As a trial solution to the equation $[F,G]=iH$ we will assume that $G$
is some diagonal matrix, with real entries.  The condition $[F,G] iH$ is then equivalent to $i H_{jk} = F_{jk}(G_{kk}-G_{jj})$.  This
suggests imposing the condition that the diagonal entries of $G$ all
be distinct, and defining
\begin{eqnarray}
  F_{jk} \equiv \left\{ \begin{array}{ll}
      \frac{i H_{jk}}{G_{kk}-G_{jj}} & \mbox{if } j \neq k; \\
      0 & \mbox{if } j = k. \\
      \end{array} \right.
\end{eqnarray}
It is easy to see that with these choices $F$ and $G$ are Hermitian,
and satisfy $[F,G] = iH$.

What of the norms $\| F \|$ and $\| G \|$?  Suppose we choose the
diagonal entries of $G$ as $-(d-1)/2,-(d-1)/2+1,\ldots,(d-1)/2$.  With
this choice it is clear from our definition that the entries of $F$
satisfy $|F_{jk}| \leq |H_{jk}|$, and we have
\begin{eqnarray}
\|F\|^2 & \leq & \mbox{tr}(F^2) \\
 & \leq & \mbox{tr}(H^2) \\
 & \leq & d \| H \|^2,
\end{eqnarray}
where the first and third inequalities follow easily from the
definitions, and the second inequality follows from the fact that
$|F_{jk}| \leq |H_{jk}|$.  With these choices we therefore have $\|F
\| \leq \sqrt{d \|H \|}$ and $\| G \| = (d-1)/2$.  Rescaling $F$ and
$G$ appropriately we can ensure that the equation $[F,G] = iH$ remains
satisfied, while satisfying
Eq.~(\ref{eq:norm_commutator_inequalities}).  This completes the
proof.

\begin{lemma} {} \label{lemma:lie-approx}
Suppose $F$ and $G$ are Hermitian matrices such
  that $\|F\|,\|G\|<\delta$.  Then:
\begin{eqnarray}
  d\left( \exp(iF) \exp(iG) \exp(-iF) \exp(-iG), \exp(i \times i[F,G])
\right)
  \leq c_1 \delta^3,
\end{eqnarray}
for some constant $c_1 \approx 4$.
\end{lemma}

\textbf{Proof:} This is easily verified using the standard series
expansion for matrix exponentials. Alternately, this is also a
standard result in the theory of Lie groups.  See, e.g., Proposition~2
on page~25 in Section~1.3 of~\cite{Rossmann02a}.

The result we desire may be obtained by combining these two lemmas.
Suppose $d(I,U) < \epsilon$.  We can find Hermitian $H$ such that $U \exp(iH)$ and $d(I,U) = \| H \| + O(\| H\|^3)$.  By
Lemma~\ref{lemma:lie-solution} we can find Hermitian $F$ and $G$ such
that $[F,G]=iH$, and $\|F \|,\|G\| \leq c_{\rm gc''} \sqrt{\epsilon}$
for $c_{\rm gc''} \approx d^{1/4} \sqrt{(d-1)/2}$.  Setting $V \equiv
\exp(iF), W \equiv \exp(iG)$, $\delta \equiv c_{\rm gc''}
\sqrt{\epsilon}$ and applying Lemma~\ref{lemma:lie-approx}, we obtain:
\begin{eqnarray}
  d(VWV^\dagger W^\dagger, U) < c_1 (c_{\rm gc''}\sqrt{\epsilon})^3
= c_{\rm gc'} \epsilon^{3/2},
\end{eqnarray}
where $c_{gc'} = c_1 c_{\rm gc ''}^3 \approx 4
d^{3/4}((d-1)/2)^{3/2}$.  Furthermore, we have $d(I,V), d(I,W) <
c_{\rm gc''} \sqrt{\epsilon}$, as desired.

\section{Prior work}
\label{sec:prior-work}

The treatment of the SK theorem and algorithm given in this paper is
based upon Appendix~3 in~\cite{Nielsen00a}.  That appendix gave a
detailed description of the SK theorem, and included an exercise
(Exercise~A3.6) asking the reader to find an algorithm for efficiently
finding accurate sequences of instructions approximating a desired
unitary.  Reader feedback on~\cite{Nielsen00a} suggests that this was
not one of the easier exercises in that volume, a fact that partially
inspired the present review paper.

The history of the SK theorem and algorithm is interesting.  For
$SU(2)$ the SK theorem was announced by Solovay in 1995 on an email
discussion list, but no paper has subsequently appeared.
Independently, in 1997 Kitaev~\cite{Kitaev97b} outlined a proof in a
review paper, for the general case of $SU(d)$.  After hearing of this
result, Solovay announced that he had generalized his proof in a
similar fashion.  Kitaev's paper also addressed the question of
efficient implementation, sketching out an algorithm to quickly find
accurate instruction sequences approximating a desired gate.  The
proof of the SK theorem given in~\cite{Nielsen00a} was based
on~\cite{Kitaev97b}, discussions with Kitaev, Solovay, and
M.~Freedman, and on a 1999 lecture by Freedman, based on Kitaev's
proof.

Kitaev's 1997 discussion~\cite{Kitaev97b} of the SK theorem and
algorithm is described and extended in the 2002 text by Kitaev, Shen
and Vyalyi~\cite{Kitaev02a}.  Indeed, that text actually presents two
different approaches to the problem.  The first approach, described in
Section~8.3 of~\cite{Kitaev02a}, uses very similar ideas to the
description we have given, with a few technical differences resulting
in somewhat different behaviour for the algorithm's running time, and
for the length of the instruction sequences produced.  In particular,
the algorithm in~\cite{Kitaev02a} has a running time
$O(\log^{3+\delta}(1/\epsilon))$ and produces an instruction sequence
of length $O(\log^{3+\delta}(1/\epsilon)$, where $\delta$ can be
chosen to be any positive real number.

The second approach, described in Section~13.7 of~\cite{Kitaev02a}, is
quite different in flavour.  It modifies our setting for the SK
theorem and algorithm in two ways: (1) it makes use of ancilla qubits,
and (2) it constrains the allowed instruction set somewhat, for
example, to Clifford group gates, together with the Toffoli gate;
certain other instruction sets are also possible, but not an arbitrary
instruction set.  A variant of Kitaev's phase estimation
algorithm~\cite{Kitaev95a,Kitaev97a} is used to construct instruction
sequences providing exponentially accurate conditional phase shifts.
Standard constructions (e.g.~\cite{Barenco95a}), together with these
phase shift gates, can then be used to approximate the desired gates.
The running time for this approach is $O(\log^2(1/\epsilon)
\log(\log(1/\epsilon)))$, and the length of the instruction sequence
is $O(\log^2(1/\epsilon) \log(\log(1/\epsilon)))$.  This approach also
has the advantage of being easily parallelizable, when acting on
multiple qubits.

A non-constructive approach to the SK theorem was taken by Harrow,
Recht and Chuang in~\cite{Harrow02a}, based on Harrow's undergraduate
thesis~\cite{Harrow01a} and papers by Arnold and
Krylov~\cite{Arnold62a}, and by Lubotsky, Phillips and
Sarnak~\cite{Lubotsky86a,Lubotsky87a} (c.f.~\cite{Gamburd99a}).  In
particular,~\cite{Harrow02a} proves that for a suitable choice of
instruction set, an approximation of accuracy $\epsilon$ may be
achieved using a sequence of $O(\log(1/\epsilon))$ instructions.  It
does not yet seem to be well understood exactly which instruction sets
achieve such a fast rate of convergence.
Furthermore,~\cite{Harrow02a} describe a simple volume argument
establishing that, up to a constant factor, it is not possible to
obtain shorter approximating sequences, and so this result is optimal.
However, they do not provide a constructive means of efficiently
finding these short instruction sequences.

We conclude by noting two papers that, while not directly concerned
with the SK theorem or algorithm, are likely of interest to anyone
interested in quantum compilation.  The first is a paper by Freedman,
Kitaev, and Lurie~\cite{Freedman02a}, who develop some general
conditions under which a subset $\mathcal{G}$ of a semisimple Lie
group $G$ must generate a dense subset of $G$.  In particular, they
introduce a natural family of metrics defined for any semisimple Lie
group $G$, and show that there is a \emph{universal} constant $c$,
independent of $G$, so that provided all points in $G$ are within a
distance $c$ of $\mathcal{G}$, then $\mathcal{G}$ must generate a
dense subset of $G$.  Roughly speaking, provided $\mathcal{G}$
``fills'' $G$ sufficiently well, it is guaranteed to generate a dense
subset of $G$.

The second paper of interest is by Fowler~\cite{Fowler04a}, who
investigates the feasibility of compiling Shor's algorithm into
fault-tolerant form using a search technique that, while while not
``efficient'' in the sense of the SK algorithm, in practice seems to
yield promising results.

\section{Conclusion}
\label{sec:conclusion}

The Solovay-Kitaev theorem and algorithm are fundamental results in
the theory of quantum computation, and it seems likely that variants
of these results will be used in future implementations of quantum
computers to compile quantum algorithms such as Shor's into a
fault-tolerant form.  The discussion of these results in the present
review paper has been pedagogical, aimed at a formulation that brings
out the key ideas, rather than being optimized for accuracy and
efficiency.  It is an interesting open problem to determine the extent
to which these constructions can be improved, perhaps even developing
a version of the Solovay-Kitaev algorithm achieving optimal or
near-optimal accuracy and efficiency.

\section*{Acknowledgments}

We thank Mike Freedman, Aram Harrow, and Alexei Kitaev for
enlightening discussions about the Solovay-Kitaev theorem and
algorithm, and to Mark de Burgh, Mark Dowling, and Yeong Cherng Liang
for helpful comments on a draft of the paper.


\begin{thebibliography}{10}

\bibitem{Arnold62a}
V.~I. Arnold and A.~L. Krylov.
\newblock {\em Soviet Math. Dokl.}, 4:1, 1962.

\bibitem{Barenco95a}
A.~Barenco, C.~H. Bennett, R.~Cleve, D.~P. DiVincenzo, N.~Margolus, P.~Shor,
  T.~Sleator, J.~A. Smolin, and H.~Weinfurter.
\newblock Elementary gates for quantum computation.
\newblock {\em Phys. Rev. A}, 52:3457--3467, 1995.
\newblock {arXiv}:quant-ph/9503016.

\bibitem{Bernstein97a}
E.~Bernstein and U.~Vazirani.
\newblock Quantum complexity theory.
\newblock {\em SIAM J. Comp.}, 26(5):1411--1473, 1997.
\newblock {arXiv}:quant-ph/9701001.

\bibitem{Fowler04a}
A.~G. Fowler.
\newblock Constructing arbitrary single-qubit fault-tolerant gates.
\newblock {\em {arXiv}:quant-ph/0411206}, 2004.

\bibitem{Freedman02a}
M.~Freedman, A.~Kitaev, and J.~Lurie.
\newblock Diameters of homogeneous spaces.
\newblock {\em {arXiv}:quant-ph/0209113}, 2002.

\bibitem{Gamburd99a}
A.~Gamburd, D.~Jakobson, and P.~Sarnak.
\newblock Spectra of elements in the group ring of $su(2)$.
\newblock {\em J. Eur. math. Soc.}, 1(1):51--85, 1999.

\bibitem{Grover96a}
Lov~K. Grover.
\newblock A fast quantum mechanical algorithm for database search.
\newblock In {\em 28th ACM Symposium on Theory of Computation}, page 212, New
  York, 1996. Association for Computing Machinery.

\bibitem{Grover97a}
Lov~K. Grover.
\newblock Quantum mechanics helps in searching for a needle in a haystack.
\newblock {\em Phys. Rev. Lett.}, 79(2):325, 1997.
\newblock {arXiv}:quant-ph/9706033.

\bibitem{Harrow02a}
A.~Harrow, B.~Recht, and I.~L. Chuang.
\newblock Efficient discrete approximations of quantum gates.
\newblock {\em J.~Math.~Phys.}, 43:4445, 2002.
\newblock {arXiv}:quant-ph/0111031.

\bibitem{Harrow01a}
A.~W. Harrow.
\newblock Quantum compiling.
\newblock MIT undergraduate thesis, 2001.

\bibitem{Horn91a}
R.~A. Horn and C.~R. Johnson.
\newblock {\em Topics in matrix analysis}.
\newblock Cambridge University Press, Cambridge, 1991.

\bibitem{Kitaev97b}
A.~Y. Kitaev.
\newblock Quantum computations: algorithms and error correction.
\newblock {\em Russ. Math. Surv.}, 52(6):1191--1249, 1997.

\bibitem{Kitaev97a}
A.~Y. Kitaev.
\newblock Quantum error correction with imperfect gates.
\newblock In A.~S.~Holevo O.~Hirota and C.~M. Caves, editors, {\em Quantum
  Communication, Computing, and Measurement}, pages 181--188, New York, 1997.
  Plenum Press.

\bibitem{Kitaev02a}
A.~Y. Kitaev, A.~H. Shen, and M.~N. Vyalyi.
\newblock {\em Classical and quantum computation}, volume~47 of {\em Graduate
  Studies in Mathematics}.
\newblock American Mathematical Society, Providence, Rhode Island, 2002.

\bibitem{Kitaev95a}
A.~Yu. Kitaev.
\newblock Quantum measurements and the \mbox{Abelian} stabilizer problem.
\newblock {\em \mbox{arXiv}:quant-ph/9511026,}, 1995.

\bibitem{Knuth74a}
D.~E. Knuth.
\newblock Computer science and its relation to mathematics.
\newblock {\em Amer. Math. Month.}, 81(4), April 1974.

\bibitem{Lubotsky86a}
A.~Lubotsky, R.~Phillips, and P.~Sarnak.
\newblock Hecke operators and distributing points on the sphere i.
\newblock {\em I.~Comm.~Pure~Appl.~Math.}, 39:S149--S186, 1986.

\bibitem{Lubotsky87a}
A.~Lubotsky, R.~Phillips, and P.~Sarnak.
\newblock Hecke operators and distributing points on the sphere ii.
\newblock {\em I.~Comm.~Pure~Appl.~Math.}, 40:401--420, 1987.

\bibitem{Nielsen00a}
M.~A. Nielsen and I.~L. Chuang.
\newblock {\em Quantum computation and quantum information}.
\newblock Cambridge University Press, Cambridge, 2000.

\bibitem{Preskill98b}
J.~Preskill.
\newblock Reliable quantum computers.
\newblock {\em Proc. Roy. Soc. A: Math., Phys. and Eng.}, 454(1969):385--410,
  1998.

\bibitem{Rossmann02a}
W.~Rossmann.
\newblock {\em Lie Groups: An Introduction Through Linear Groups}.
\newblock Oxford University Press, Oxford, 2002.

\bibitem{Shor94a}
P.~W. Shor.
\newblock Algorithms for quantum computation: discrete logarithms and
  factoring.
\newblock In {\em Proceedings, 35th Annual Symposium on Fundamentals of
  Computer Science}, Los Alamitos, 1994. IEEE Press.

\bibitem{Shor97a}
P.~W. Shor.
\newblock Polynomial-time algorithms for prime factorization and discrete
  logarithms on a quantum computer.
\newblock {\em SIAM J. Comp.}, 26(5):1484--1509, 1997.

\end{thebibliography}

\end{document}